\documentclass[aps,prb,reprint,superscriptaddress]{revtex4-2}
\usepackage{amsmath}
\usepackage{amssymb}
\usepackage{graphicx}
\usepackage{hyperref}
\usepackage{color,xcolor}
\begin{document}
\title{Ferroelectric control of spin-polarized two-dimensional electron gas}
\author{Yakui Weng}
\email{Corresponding author: wyk@njupt.edu.cn}
\affiliation{School of Science, Nanjing University of Posts and Telecommunications, Nanjing 210023, China}
\author{Wei Niu}
\affiliation{School of Science, Nanjing University of Posts and Telecommunications, Nanjing 210023, China}
\affiliation{Laboratory of Solid State Microstructures Nanjing University, Nanjing 210093, China}
\author{Xin Huang}
\affiliation{School of Science, Nanjing University of Posts and Telecommunications, Nanjing 210023, China}
\author{Ming An}
\author{Shuai Dong}
\affiliation{School of Physics, Southeast University, Nanjing 211189, China}
\date{\today}

\begin{abstract}
Spin-polarized two-dimensional electron gas ($2$DEG) at oxide interfaces is an emerging physical phenomenon, which is technologically important for potential device applications. However, most previous relevant studies only focused on the creation and characterization of the spin-polarized $2$DEG. To push forward the device applications, the control of spin-polarized $2$DEG by electric field is an important step. Here, a model system based on antiferromagnetic and ferroelectric perovskites, i.e., YTiO$_3$/PbTiO$_3$ superlattice, is designed to manipulate the spin-polarized $2$DEG. By switching the direction of polarization, the spin-polarized $2$DEG can be effectively tuned for both symmetric interfaces and asymmetric polar interfaces.
\end{abstract}
\maketitle

\section{Introduction}
As one of the most intriguing physical phenomena of electronic reconstruction, the metallic interface with high carrier mobility between two insulating oxides, i.e., two-dimensional electron gas ($2$DEG), provides a unique platform for exploring fundamental physics and electronic devices \cite{Ohtomo:Nat,Ohtomo:Nat04,Nakagawa:Nm,Caviglia:Nat,Bhalla:Np}. Comparing with conventional $2$DEG in semiconductor quantum wells which are formed by the $s$ or $p$ orbital electrons, the $2$DEG at complex oxide interface originated from the $d$ orbitals electrons own more degrees of freedoms \cite{Gabay:Np,Bristowe:Jpcm14}, especially the spin, which leads to the magnetism.

However, most previous works of $2$DEG at oxide interfaces were based on SrTiO$_3$, such as LaAlO$_3$/SrTiO$_3$ (LAO/STO) \cite{Li:Np11,Brinkman:Nm07,Dikin:Prl11,Bert:Np11,Smink:Prl17} and $\gamma$-Al$_2$O$_3$/STO \cite{Niu:Nl17,Chen:Am14,Cao:Npjqm16,Christensen:Np19}, in which the interfacial magnetism is very weak since their parent materials are nonmagnetic. To enhance the interfacial magnetism, systems with magnetic insulators were studied to support spin-polarized $2$DEGs \cite{Wang:Prb09,Moetakef:Prx12,Betancourt:Prb17,Lomker:Prm17,Zhang:Prl18,Zhang:Nl19}. For example, high mobility spin-polarized $2$DEGs have been obtained by growing ferromagnetic (FM) EuO on KTaO$_3$ \cite{Zhang:Prl18}, while the calculated magnetic moments of interfacial Ta atom ($\sim0.18$ $\mu_{\rm B}$/Ta) still need to be improved.

A more important issue for spin-polarized $2$DEG is how to manipulate it using electric field, i.e., a converse magnetoelectric (ME) effect. However, reports on this issue are scarce. Although a switchable $2$DEG at ferroelectric (FE) interfaces has been predicted in a few systems, such as PbTiO$_3$/STO or symmetric KNbO$_3$/$A$TiO$_3$ ($A$=Sr, Ba, Pb) \cite{Niranjan:Prl09,Puente:Prb15}, the spin polarization has not been mentioned or emphasized due to the nonmagnetic nature of their parent materials. Even in previous studies of spin-polarized $2$DEG containing FM material, it is not easy to obtain spin-dependent switching function. Since in the presence of robust FM order, the direction of induced spin polarization at the interface is correspondingly stable and thus difficult to manipulate.

\begin{figure}[h!]
\centering
\includegraphics[width=0.46\textwidth]{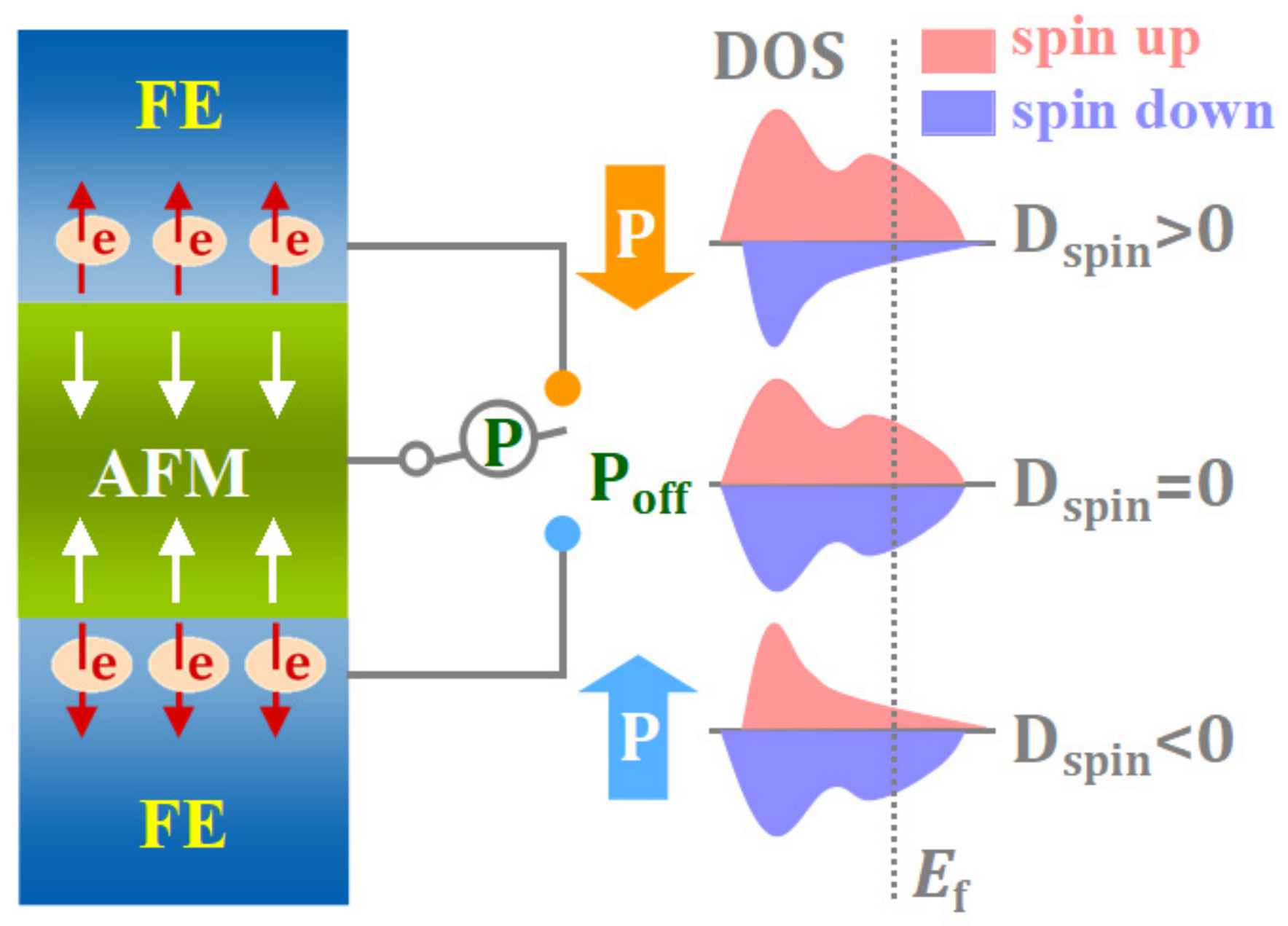}
\caption{Schematic function of spin-dependent switching effect. Narrow arrows (red and white) and wide arrows (orange and blue) denote the spin and polarization directions, respectively. Electrons are attracted to the interfaces under ferroelectric field effect. Considering the interfacial AFM coupling, the sign of interfacial spin density (i.e. spin-polarized $2$DEG) is turned accompanying the switch of polarization ($\textbf{P}$). Even without an applied electric field, $2$DEGs still exists.}
\label{F1}
\end{figure}

To realize this spin-dependent switching effect [shown in Fig.~\ref{F1}], a model system based on antiferromagnetic and ferroelectric perovskites is proposed here to manipulate the spin-polarized $2$DEG. The key suggestion is to replace FM material by A-type antiferromagnetic (AFM) material. Because both the FE field effect ($\nabla\cdot\textbf{P}$) and A-type AFM order are layer dependent \cite{Dong:Prb13.2,Weng:Prl16}, A-type AFM order will be better coupled with the field effect. In this case, the above mentioned switching function can be achieved by adjusting the interfacial spin density. The scheme of this design is shown in Fig.~\ref{F1}. Within the interfacial AFM coupling, the spin-polarized $2$DEG can be naturally tuned by electric field. Even without electric field, $2$DEG still exists.

For a proof of the concept, YTiO$_3$/PbTiO$_3$ (YTO/PTO) interface along the [$001$] direction will be studied as a model system. PTO is one of the most studied ferroelectric oxides with large polarization $\sim75-80$ $\mu$C/cm$^{2}$ \cite{Gavrilyachenko:Spss,Haun:Jap,Sharma:Prb,Tang:Sci,zhang:Fp15,Damodaran:Am17,Lu:Prl18}. Altough bulk YTO is FM, the YTO film grown on LAO substrate can easily become A-type AFM order \cite{Huang:Jap13}. Based on density functional calculations, we find that the spin-polarized $2$DEGs can be formed between AFM insulator and ferroelectric oxide. Upon the FE switching, the corresponding spin-polarization shows significant modulations, a desired function of spintronics.

\section{Model \& method}
PTO is a $d^0$ band insulator with large band gap of $\sim3.4$ eV. At room temperature, it has a tetragonal structure (space group $P4mm$) with lattice constants of $a$=$b$=$3.905$ {\AA} and $c$=$4.156$ {\AA}, giving a moderate tetragonality ($c/a$=$1.064$) \cite{Glazer:Acb2}. The ground state of YTO bulk is a FM Mott insulator with GdFeO$_3$-type distortion \cite{Mochizuki:Njp}. The space group is $Pbnm$ and the lattice constants are $a$=$5.338$ {\AA}, $b$=$5.690$ {\AA}, and $c$=$7.613$ {\AA} \cite{Goral:Jssc}, as sketched in Fig.~\ref{F2}(a). In the following, the YTO/PTO superlattices are assumed to be grown on the widely used LAO ($001$) substrate. To match the substrate, the in-plane lattice constants of YTO and PTO are fixed as $3.794\times\sqrt{2}$=$5.366$ {\AA}.

DFT calculations were performed based on the generalized gradient approximation (GGA) with Perdew-Burke-Ernzerhof (PBE) potentials, as implemented in the Vienna $ab$ $initio$ simulation package (VASP) code \cite{Blochl:Prb,Kresse:Prb99}. The cutoff energy of plane-wave is $520$ eV. The Hubbard repulsion $U_{\rm eff}$=$U-J$ is imposed on Ti's $3d$ orbitals using the Dudarev implementation \cite{Dudarev:Prb}. According to previous literature \cite{Zhang:Prb15}, $U_{\rm eff}$(Ti)=$3.2$ eV is proper to reproduce the experimental properties and thus is adopted as default parameters in the following calculations. Monkhorst-Pack \textit{k}-point meshes of $9\times9\times1$ centered at $\varGamma$ point adopted for YTO/PTO superlattices stacking along the [$001$] direction. Both the out-of-plane lattice constants and atomic positions are fully relaxed until the Hellman-Feynman forces are converged to less than $0.01$ eV/{\AA}.

\section{Results \& discussion}
First, the parent materials have been checked. Starting from the experimental structure, the lattice constants and atomic positions are fully relaxed. To obtain the magnetic ground state of YTO, the total energies of A-type, C-type, and G-type AFM and FM states are calculated. Our calculation confirms that the FM order has the lowest energy. The calculated band gap and local magnetic moment are $1.6$ eV and $0.89\mu_{\rm B}$/per Ti, respectively, slightly larger than experimental values ($1.2$ eV and $0.84\mu_{\rm B}$/per Ti) \cite{Okimoto:Prb,Garrett:Mrb}. For PTO, the calculated polarization is $86.6$ $\mu$C/cm$^{2}$, which is close to the experimental value and previous theoretical value \cite{Gavrilyachenko:Spss,zhang:Fp15}. All these results guarantee the reliability of following calculations on superlattices.

\begin{figure}
\centering
\includegraphics[width=0.46\textwidth]{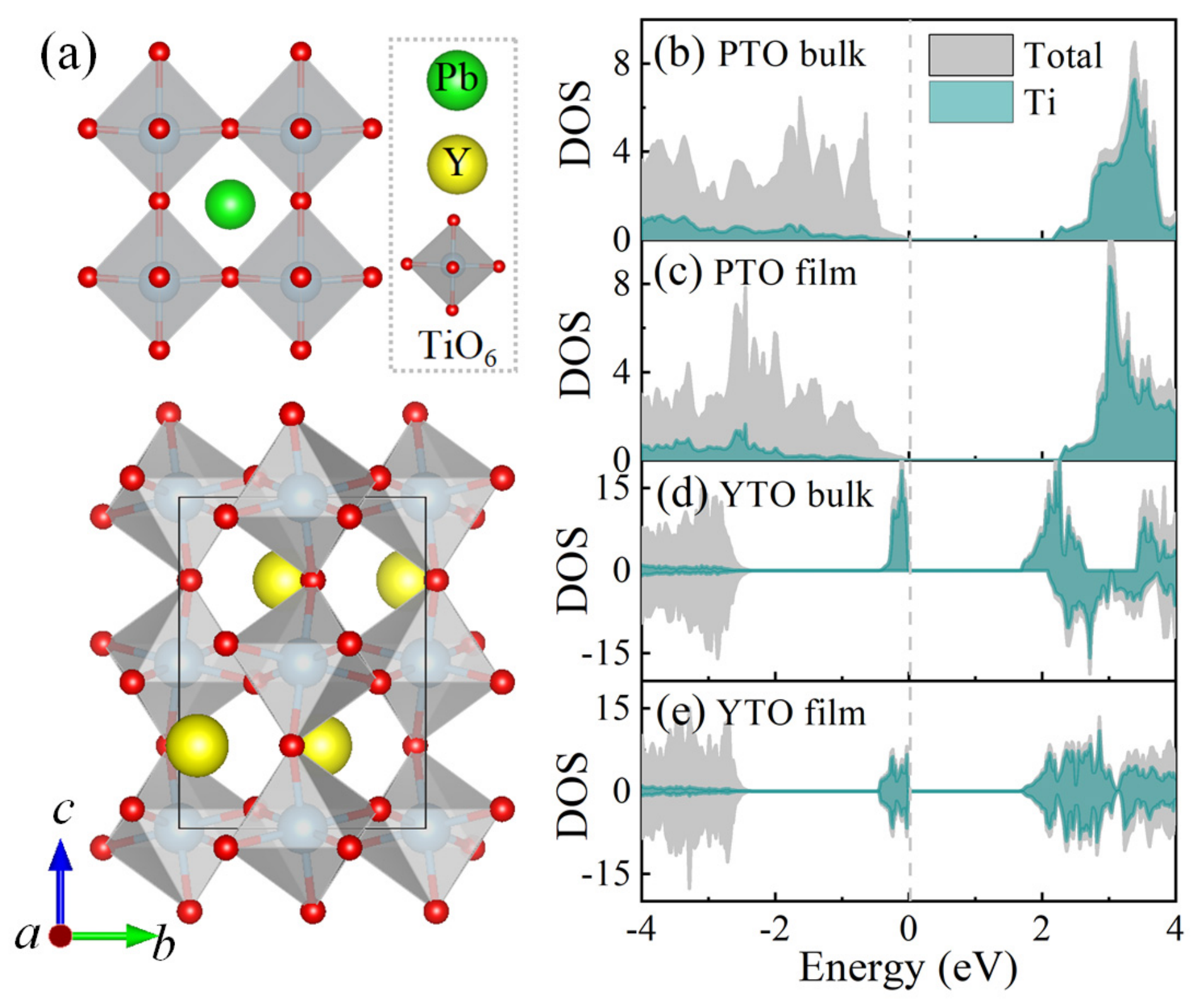}
\caption{(a) Schematic crystal structures of PTO and YTO. (b)-(e) Projected density of states (PDOS) for bulks and films: (b) PTO bulk, (c) PTO film, (d) YTO bulk, (e) YTO film. The Fermi level for each case is set as zero and marked by (gray) broken line.}
\label{F2}
\end{figure}

Then the strain effects from LAO substrate have been studied. The total density of states (DOS) and atomic projected density of states (PDOS) of bulks and films are displayed in Fig.~\ref{F2}(b)-(e). Upon the compressive strain, PTO film remains insulating as in the unstrained conditions, as shown in Fig.~\ref{F2}(b) and (c). Similarly, the insulating behavior and the band gap of YTO are also not significantly affected by this strain [Fig.~\ref{F2}(d) and (e)]. However, due to the lattice distortions, YTO film undergoes a phase transition from the FM state to A-type AFM state, further confirmed by the PDOS, in agreement with previous study \cite{Huang:Jap13}.

For YTO/PTO superlattices, the magnetic and electronic structures of both symmetric and asymmetric interfaces are studied, as shown in Fig.~\ref{F3}(a) and Fig.~\ref{F4}(a). To keep the PTO bulk polarization in positive or negative direction, PTO layers are fixed, while the lattice length along the $c$ axis and the atomic positions of YTO are relaxed. In addition, the calculations with optimized interfacial three layers of PTO (namely, TiO$_2$/PbO/TiO$_2$) are also tested for comparison, as shown in SM \cite{Supp2}, which do not alter the physical conclusion.

\subsection{(YTO)$_{2.5}$/(PTO)$_{5.5}$ superlattice}
\begin{figure}
\centering
\includegraphics[width=0.46\textwidth]{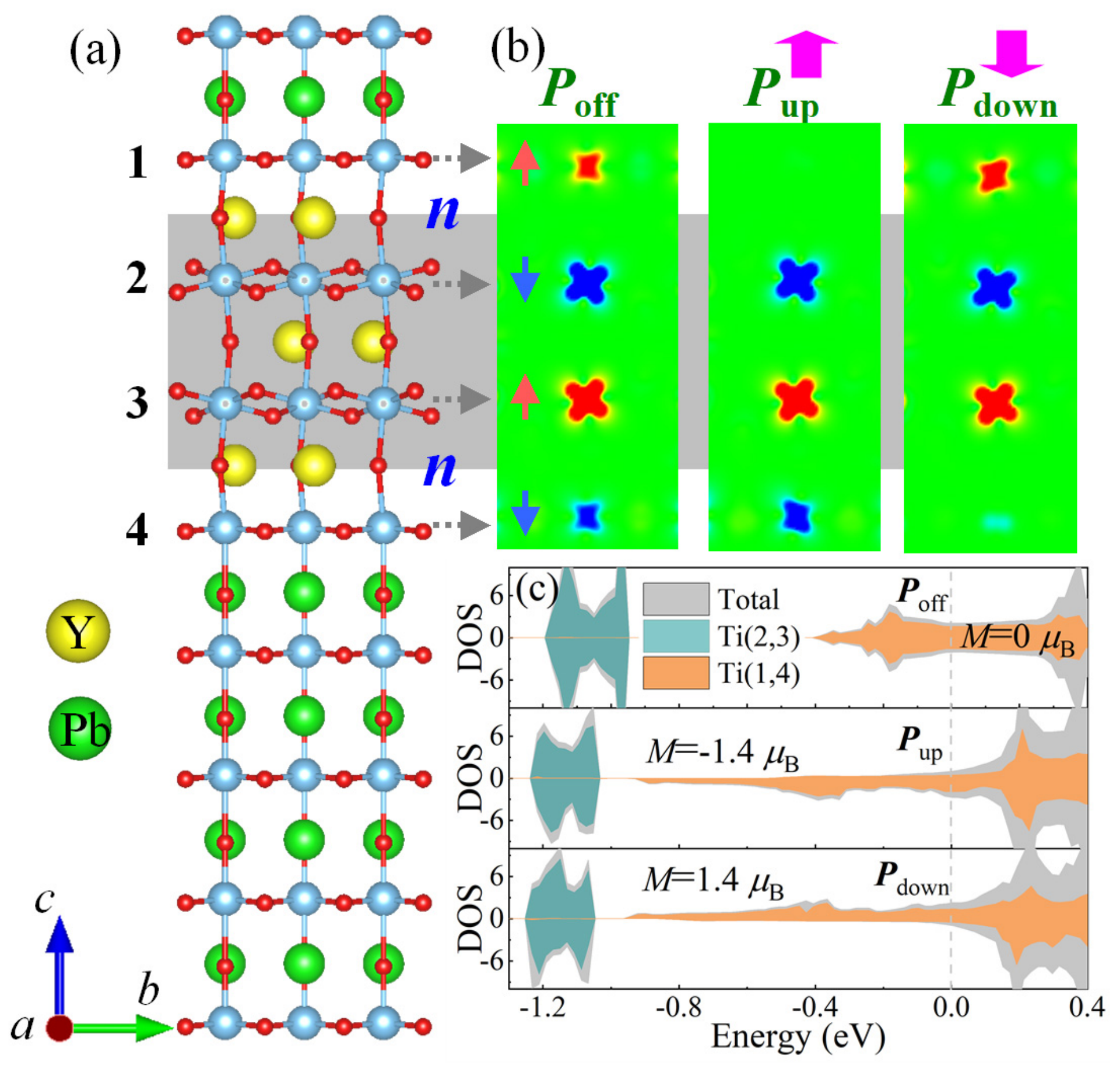}
\caption{(a) Crystalline structure of (YTO)$_{2.5}$/(PTO)$_{5.5}$ superlattice ($80$ atoms) grown along the [$001$] direction. The lattice structure is periodic without vacuum layer. $1$, $2$, $3$ and $4$ stand for the TiO$_2$ layers at the interface. The $n$-type interfaces are indicated. (b) Spatial distribution of the spin density for the cases (left) $\textbf{P}_{\rm off}$ (without ferroelectric polarization), (middle) $\textbf{P}_{\rm up}$ and (right) $\textbf{P}_{\rm down}$. The spins are distinguished by colors. (c) The corresponding DOS and PDOS. The Fermi level for each case is set as zero and marked by (gray) broken line. The total magnetization $\textbf{M}$ are indicated.}
\label{F3}
\end{figure}

(YTO)$_{2.5}$/(PTO)$_{5.5}$ superlattice stacked along the [$001$] axis with symmetric interfaces is studied first, as shown in Fig.~\ref{F3}(a). Here three layers of (YO)$^{+1}$ with TiO$_2$-YO-TiO$_2$ interfaces (i.e. the double $n$-type interfaces) are adopted. In this scenario, one more electron is introduced into the system due to the uncompensated ionic charge on the additional (YO)$^{+1}$ layer. As the first step, the magnetic ground state is checked. With the optimized $c$-axis lattice constant, the A-type AFM state has a lower energy than FM one, similar to the YTO film, as expected.

In this superlattice with Y trilayer, the TiO$_2$ layers at the interfaces and inside YTO are labeled as ($1$, $4$) and ($2$, $3$), respectively, as shown in Fig.~\ref{F3}(a). Under the FE field effect, excess electrons are attracted to the interfaces. Then, by virtue of the restriction of layered AFM coupling between Ti layers, the local magnetic moments of the interfacial Ti ions show significant modulations accompanying the switch of $\textbf{P}$.

As shown in Fig.~\ref{F3}(b) and Table~\ref{table1}, without the FE $\textbf{P}$ (i.e. $\textbf{P}_{\rm off}$), extra electrons are equally distributed between $1$st and $4$th Ti layers due to the symmetric interfaces, which gives rise to a $0$ $\mu_{\rm B}$ net moment. When the FE $\textbf{P}$ is parallel to the $c$ axis (i.e. $\textbf{P}_{\rm up}$), extra electrons are collected on the $4$st Ti layer, and the spin direction is opposite to the $3$nd Ti layer due to AFM coupling. The calculated local magnetic moment of interfacial Ti is $\mathbin{\sim}-0.64$ $\mu_{\rm B}$/Ti, leading to a net magnetization $\textbf{M}\mathbin{\sim}-1.4$ $\mu_{\rm B}$ (two Ti ions per layer), larger than the EuO/KTaO$_3$ one ($\sim0.18$ $\mu_{\rm B}$/Ta). Similarly, when the FE $\textbf{P}$ is antiparallel to the $c$ axis (i.e. $\textbf{P}_{\rm down}$), electrons are collected on the $1$st Ti layer, giving a net magnetization $\textbf{M}\mathbin{\sim}+1.4$ $\mu_{\rm B}$. Therefore, the sign of $\textbf{M}$ is switchable upon electric switching.

\begin{table}
\caption{Local magnetic moments for the case of symmetric and asymmetric interfaces. $m_1$, $m_2$ $m_3$ and $m_4$ are the local magnetic moments for the $1$st, $2$nd, $3$rd and $4$th TiO$_2$ layers, respectively, integrated within the Wigner-Seitz spheres. All moments are in units of $\mu_{\rm B}$.}
\begin{tabular*}{0.48\textwidth}{@{\extracolsep{\fill}}llcccc}
\hline \hline
Superlattice & FE & $m_1$ & $m_2$ & $m_3$ & $m_4$\\
\hline
& $\textbf{P}_{\rm up}$ & $0$ & $-0.895$ & $0.892$ & $-0.639$\\
(YTO)$_{2.5}$/(PTO)$_{5.5}$ & $\textbf{P}_{\rm off}$ & $0.424$ & $-0.889$ & $0.889$ & $-0.424$\\
& $\textbf{P}_{\rm down}$ & $0.643$ & $-0.892$ & $0.895$ & $0$\\
\hline
& $\textbf{P}_{\rm up}$ & $0$ & $-0.407$ & $0.879$ & $-0.524$\\
(YTO)$_{2}$/(PTO)$_{6}$ & $\textbf{P}_{\rm off}$ & $0$ & $-0.812$ & $0.855$ & $0$\\
& $\textbf{P}_{\rm down}$ & $0$ & $-0.878$ & $0.862$ & $0$\\
\hline \hline
\end{tabular*}
\label{table1}
\end{table}

By studying the DOS in Fig.~\ref{F3}(c), we found that the system presents metallic behavior in all cases. Considering the insulating YTO and PTO with relatively large band gaps, the metal state (conduction charge) is believed to originate from the interfacial charge reconstruction. In addition, the states at the Fermi level are mainly contributed by the interfacial Ti layers of PTO, implying the interfacial $2$DEG. Therefore, the coexistence of magnetism $2$DEG, i.e., spin-polarized $2$DEG, is presented. Furthermore, this spin-polarized $2$DEG can be switched accompanying the flipping of $\textbf{P}$, a desired spin-dependent switching function.

In fact, for such magnetoelectric system with polarization and antiferromagnetism, the field-effect ME coupling can be expressed as ($\nabla\cdot\textbf{P}$)($\textbf{M}\cdot\textbf{L}$), as in BiFeO$_3$/SrTiO$_3$ heterostructure \cite{Weng:Prl16}, where $\textbf{L}$ is the AFM order parameter. Phenomenologically, under this ME energy term, when magnetization and polarization are switched together, it is equivalent to rotating the crystal structure along the axis for symmetric interfaces. In this sense, the results presented in Fig.~\ref{F3} and Table~\ref{table1} are expectable.

\subsection{(YTO)$_{2}$/(PTO)$_{6}$ superlattice}
\begin{figure}
\centering
\includegraphics[width=0.46\textwidth]{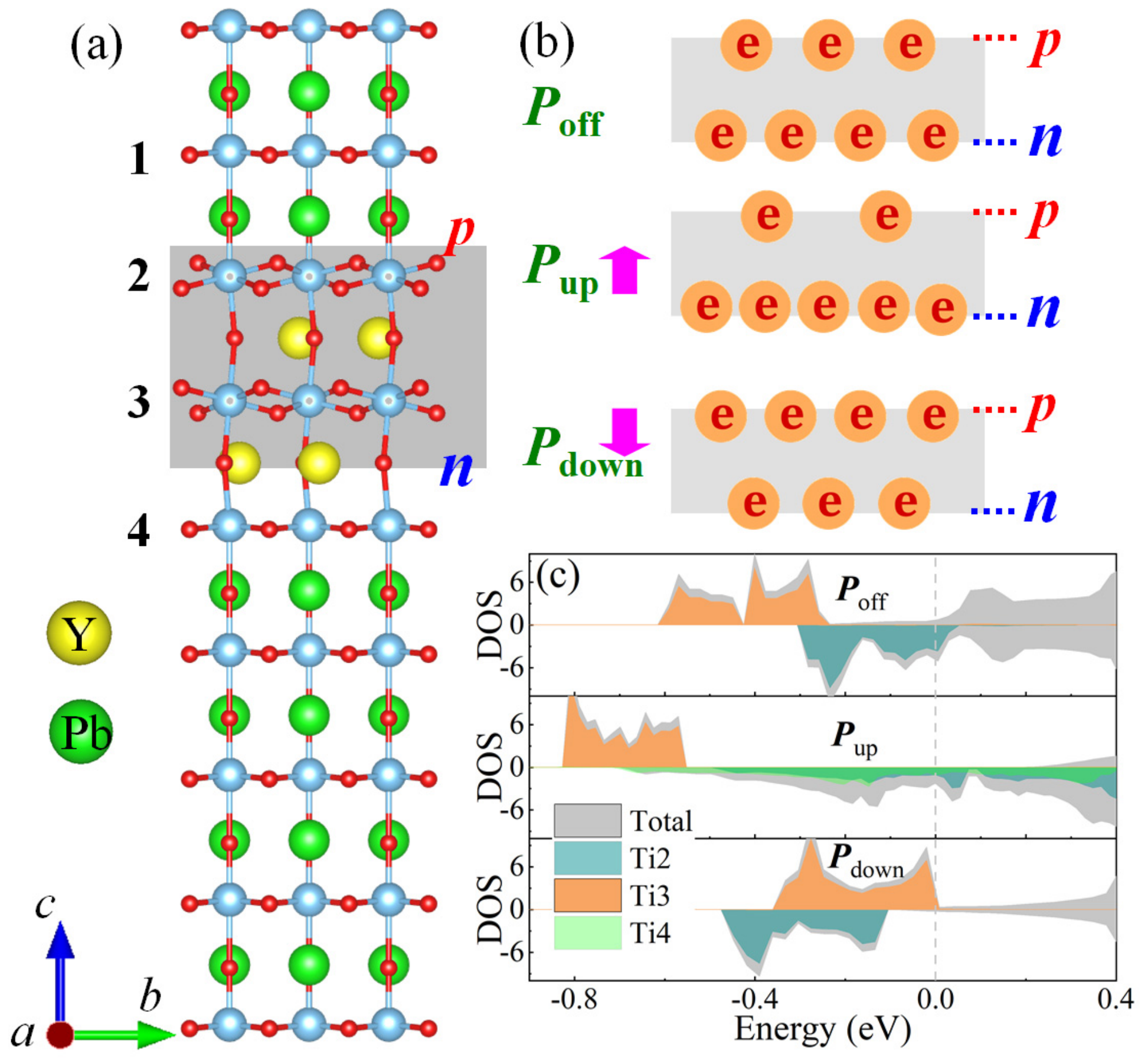}
\caption{(a) Crystalline structure of (YTO)$_{2}$/(PTO)$_{6}$ superlattice ($80$ atoms) grown along the [$001$] direction. The lattice structure is periodic without vacuum layer. $1$, $2$, $3$ and $4$ stand for the TiO$_2$ layers at the interface. The $n$-/$p$-type interfaces are indicated. (b) The interfacial electronic density modulated by asymmetric interfaces and FE $P$ (pink arrows). (c) The corresponding DOS and PDOS. The Fermi level for each case is set as zero and marked by (gray) broken line.}
\label{F4}
\end{figure}

Subsequently, the calculation is done for the superlattice based on the asymmetric polar interfaces. As shown in Fig.~\ref{F4}(a), the interfaces with TiO$_2$-YO-TiO$_2$ and TiO$_2$-PbO-TiO$_2$ are selected as $n$-type and $p$-type interfaces, respectively. Phenomenologically, the $n$-type interface will attract electrons to the interface, and conversely, the $p$-type interface will repel electrons away from the interface. Therefore, the interfacial charge disproportion can be naturally induced by asymmetric interfaces. Even without FE $\textbf{P}$ (i.e., $\textbf{P}_{\rm off}$), the electronic density and electrostatic potential (see Fig.~\ref{F4}(b)) are already modulated, completely different from the results in symmetric interfaces.

Then, let us discuss the situation with an applied electric field. As sketched in Fig.~\ref{F4}(b), when $\textbf{P}$'s pointing perpendicular to the $n$-type interface (i.e., $\textbf{P}_{\rm up}$), the initial electrostatic potential difference between the $2$nd and $3$rd TiO$_2$ layers will be further enlarged, thus enhancing the charge disproportion. However, when the $\textbf{P}$'s pointing perpendicular to the $p$-type interface (i.e., $\textbf{P}_{\rm down}$), the initial electrostatic potential difference from the polar interfaces will be reduced, thus suppressing the charge disproportion. As a consequence, the interfacial spin polarization, which is closely related to the carrier density, can be effectively modulated accompanying the $\textbf{P}_{\rm up}$ to $\textbf{P}_{\rm down}$ switching.

The above processes are confirmed by the atomic-projected density of states (PDOS) and local magnetic moments. As shown in Fig.~\ref{F4}(c) and Table~\ref{table1}, without FE $\textbf{P}$ (i.e. $\textbf{P}_{\rm off}$), the electron concentration of the $p$-type interface ($2$nd atomic layer) is lower than that of the $n$-type interface ($3$rd atomic layer) due to asymmetric polar interfaces, leading to the charge disproportion. Such charge disproportion makes the spin-down channel partially occupied, resulting in the spin-polarized $2$DEG. For $\textbf{P}_{\rm up}$ case, the electrostatic potential is the superposition of the asymmetric polar potential from the YTO and the FE potential from the PTO, and the original charge disproportion is further enhanced. In this case, the excessively accumulated charge at $n$-type interface will be transferred from the $3$rd atomic layer to the $4$th atomic layer, thus the quantum kinetic energy makes the superlattice metallic. According to the PDOS, the states around the Fermi level mainly come from Ti's $3d$ orbitals of the $2$nd (interfacial layer of YTO) and $4$th (interfacial layer of PTO) atomic layers.

However, for $\textbf{P}_{\rm down}$ case, the charge accumulation near the interfaces depends on the competition between asymmetric polar interfaces and FE $\textbf{P}$. If these two effects could be balanced, both the electrostatic potential and electronic distribution would become uniform. As summarized in Table~\ref{table1}, the local magnetic moments of Ti ions between $2$nd and $3$rd atomic layers are almost equal, suggesting nearly full compensation between these two effects, which is also confirmed by PDOS, obviously different from results in $\textbf{P}_{\rm off}$ and $\textbf{P}_{\rm up}$ cases. Therefore, the spin-polarized $2$DEG regulation mentioned above can be extended to asymmetric interfaces.

Finally, it should be noted that although both symmetric and asymmetric interfaces can realize the control of spin-polarized $2$DEG, the results involving $2$DEG and magnetism revealed here are not the same, indicating an unusual interface-dependent polarization control. In addition, although only a few YTO layers are studied here, the ME function can also be effective in thicker YTO layers, since the ME coupling is an interface effect \cite{Duan:Prl2,Weng:Prl16} and the inner YTO layers will not contribute to magnetization.

\section{Conclusion}
In summary, using the first-principles calculation, a model system based on antiferromagnetic and ferroelectric perovskites is proposed to pursue the controllable spin-polarized $2$DEG. For both symmetric interfaces and asymmetric polar interfaces, the combination of FE polarization and antiferromagnetism can effectively tune the spin-polarized $2$DEG accompanying the ferroelectric switching. Although the titanium oxides are studied here, the design principle is general and can be extended to other systems with polarization and antiferromagnetism. The present findings suggest a new efficient approach for spin-based information control.

\begin{acknowledgments}
This work was supported by the National Natural Science Foundation of China (Grant Nos. 11804168, 11904174, 11804165, 11834002 and 11674055), the Natural Science Foundation of Jiangsu Province (Grant Nos. BK20180736 and BK20190729), NUPTSF (Grant Nos. NY219026 and NY219024), the Natural Science Foundation of the Jiangsu Higher Education Institutions of China (19KJB510047), the Innovation Research Project of Jiangsu Province.
\end{acknowledgments}

\bibliographystyle{apsrev4-1}
\bibliography{ref}
\end{document}